# Evidence of a multi resonant system within solar periodic activity

M. A. Vukcevic, M.Sc

To demonstrate a degree of correlation between periodicity in the solar activity and a multi-resonant system three examples are considered. Although only two base frequencies were employed a relatively close correlation was obtained for periodicity, amplitude's envelope and some well-known longer-term anomalies, by using a mathematical formula: $Y = \Sigma \cos 2\pi (t-T_0)/P$.

Most of the natural events fall within one of the following categories: random, periodic or combination of these two. Solar activity definitively displays short-term randomness as well as a certain, however imperfect, longer-term periodicity. The complexities of the forces active within the solar interior have prevented formulation of an elegant scientific theory describing the process.

The solar periodic activity as represented by the sunspot number, could be considered as a response function of a multi resonant system operating within the solar mass. From charts of the existing sunspot records (Fig.1), at first glance, two basic frequencies are obvious with periods of approximately 11 and 100-110 years. However, a combination of two, much closer frequencies with periods of approximately 19.9 and 23.7 years as in the following mathematical equation:

$$Y = A \, \text{abs} \left[ \cos 2\pi (t-T_0)/P_1 + \cos (2\pi/3 + 2\pi (t-T_0)/P_2) \right]$$

produces a remarkably similar result as it is demonstrated in the diagram shown in Fig.1.

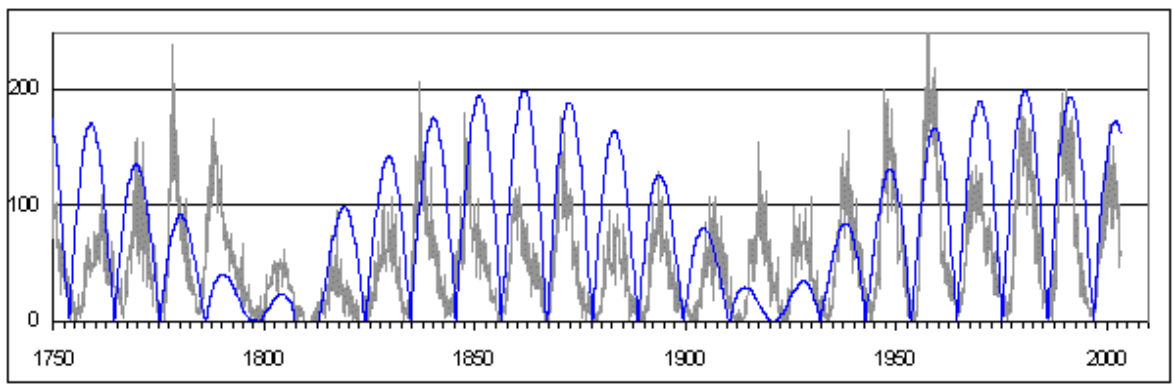

**Fig. 1**

The parameters used have following values:
- A = 100 - arbitrary constant used to normalise values in the diagram;
- P1 = 19.86 & P2 = 23.724 (in years)
- $T_0$ = 1941 – empirically chosen constant for the best coincidence, significance of which might be explained by a further research. Prior to 1813 a $90^0$ phase shift is required ( 'Sin' instead of 'Cos' functions).

To reproduce the above graph the following MS-Excel entry can be used:
=100*ABS(COS(2*PI()*(Ax-1941)/19.859)+COS(2*PI()/3+2*PI()*(Ax-1941)/23.724)).
where Ax is the A column filed containing the incremental year number.

It appears that there is a resonant system in operation within the solar body, the imperfection of which is demonstrated by the irregularity of its response. This in turn means that it should have a rich spectrum of sub-harmonics and higher harmonics. In this article attention will be focused on the sub-harmonics. There is a certain amount of justifiable scepticism about using sub-harmonics. These are not necessarily just products of beating between two or more frequencies. It could be considered that a sub-harmonic represents not only a mathematical factor, but it has a physical reality as a block or even a combination of oscillations that make-up a specific much longer cycle with distinct properties of its own.

Examining sub-harmonics of the higher order, for the above-mentioned frequencies, produces further evidence of possible existence of a multi resonant system. Multiples and sums of two periods used previously, give a number of sub-harmonics with periods that average at around 96.5 and 118 years, (P1 and P2 respectively). With these parameters and $T_0$ as above the equation:

$$Y = abs [Cos\ 2\pi\ (t-T_0)/P_1 + Cos\ 2\pi\ (t-T_0)/P_2]$$

has a following graphic representation:

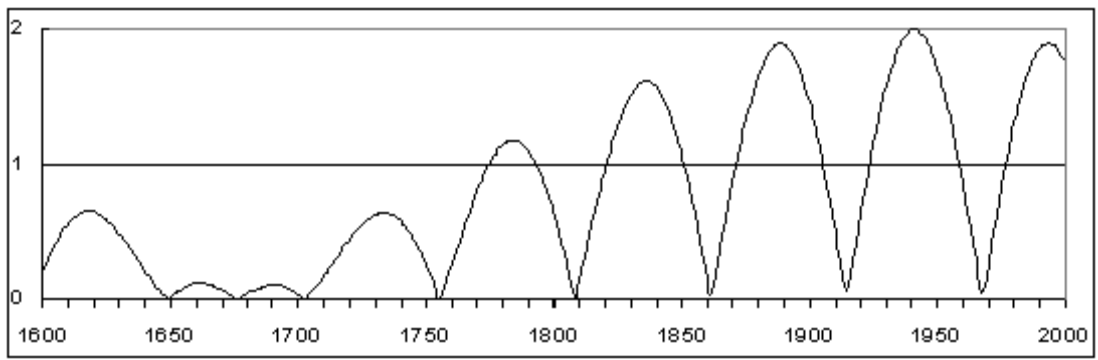

**Fig. 2**

The graph depicts anomalies within solar periodic activity with an immediately recognisable minimum between 1650-1700, coinciding with the Maunder Minimum. Further relevant dates are at or near the equation's zero value:
1809 - Dalton minimum;
1913 - another minimum but not so pronounced;
1860 and 1969 are the years of two cycles with suppressed amplitudes, 30% - 50% lower than in the neighbouring peaks, (see Fig.1.).

MS-Excel entry is: =ABS(COS(2*PI()*(Ax-1941)/96.5)+COS(2*PI()*(Ax-1941)/118))

An approximation for the overall amplitude's envelope, for the period 1800-2000, can be obtained by using sub-harmonics with periods of: P1 = 118 (as above) and P2 = 3 x 96.5 = 289.5 years and again $T_0$ = 1941. The equation is:

$$Y = A [B + Cos\ (3\pi/2 + 2\pi\ (t-T_0)/P_1 + 0.5 Cos\ 2\pi\ (t-T_0)/P_2]$$

A=60 and B=2 are the normalising coefficients.
Diagram is plotted against sunspot number graph and shown in the Fig.3.

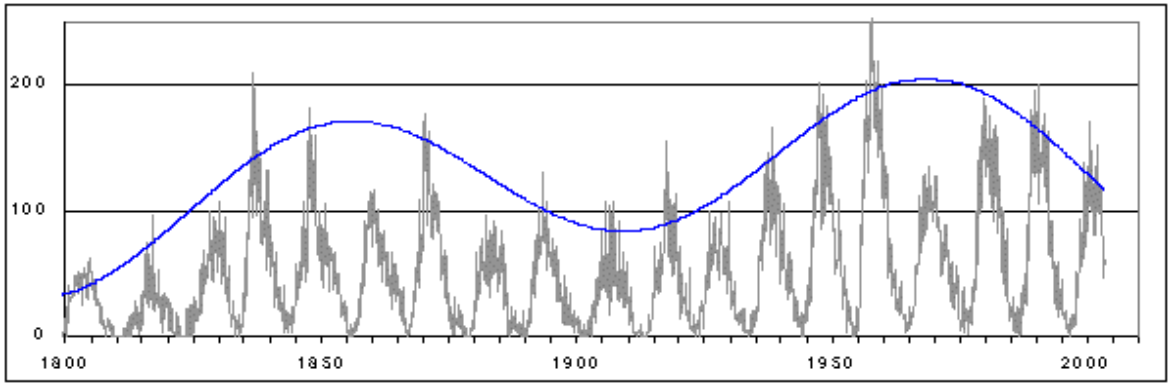

**Fig. 3**

The deviations for the cycles around the years 1860 and 1970 are explained beforehand. Therefore, only one cycle (with maximum in 1885) is significantly out. For the period prior to 1800 correlation fails.

For MS-Excel enter: =60*(2+COS(3*PI()/2+2*PI()*(Ax-1941)/118)+0.5*COS(2*PI()*(Ax-1941)/289.5))

## Conclusion

The above examples demonstrate a degree of correlation between the periodicity of the solar activity and a multi resonant system. Although only two base frequencies were considered it is possible that much closer mathematical approximation could be obtained by considering further components using the above equation in its more general form:

$$Y = \Sigma \cos 2\pi (t-T_0)/P$$

If there is indeed a relationship between solar activity and the multi resonant system it can not be absolutely and conclusively proved. It should be a matter of further scientific consideration whether to reject or accept the validity of the above demonstration.

**Reference:**
RWC Belgium, World Data Center for the Sunspot Index  http://sidc.oma.be/index.php3
Comments to: (vukcevic@ntlworld.com

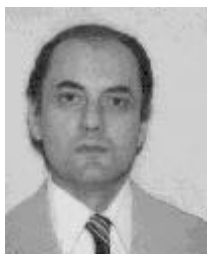

The author is
Graduate (Dipl.Ing) from Belgrade University, Yugoslavia and
Postgraduate (M.Sc) from London University, United Kingdom.